# Phase-Shifted Bell States

J.J. Joshua Davis[1], Carey L. Jackman[1], Rainer Leonhardt[1], Paul J. Werbos[2] and Maarten D. Hoogerland[1*]

[1]*Dodd-Walls Centre for Photonics and Quantum Technologies, Department of Physics, University of Auckland, Private Bag 92019, Auckland 1142, New Zealand.*

[2]*Department of Electrical & Computer Engineering, Missouri University of Science & Technology, Rolla, MO 65409, U.S.A.*

\* *m.hoogerland@auckland.ac.nz*

**Abstract:** Inspired by previous studies and pioneers of the field, we present new results on an extensive EPR-Bell experiment using photons generated by parametric down conversion, where one of the photons is deliberately phase-shifted. Our experiments show some surprising results for particular angles of this phase shift.

1. **Introduction**

The thought experiment proposed by Einstein, Podolsky and Rosen (EPR) in 1935 [1] and the subsequent proposal by Bell [2] continue to challenge the world view of students around the globe. When measuring the spin of entangled particles, EPR claimed that the 'Copenhagen' interpretation of quantum mechanics would lead to 'spooky action at a distance' and would therefore be untenable [1] [3]. EPR suggested that there must be 'hidden variables' that pre-determine the outcome of all measurements. In 1966, John Bell demonstrated that for any such theory to be correct, certain equalities [2] would need to hold. Since then, traditional sets of EPR-Bell states have been thoroughly studied [4], and even though Quantum Key Distribution (QKD) is now a commercial product that uses the concepts introduced in these papers, physicists continue to be confused and still open to inquiry in new experimental research about the nature of measurement and what is 'real' versus practical and secure, concerning classical and quantum computations and communication, when producing a reliably safe cryptographic key [5] [6].

Most of the successful experiments applying Type II spontaneous parametric down conversion (SPDC) have shown to reliably produce entangled photons [7]. In any photon pair produced this way, one is horizontally polarized and the other vertically. This makes this technique very attractive to further experiment with EPR-Bell States [8] as studied by [7] [9] [10].

In this paper, we present an extensive range of experiments on the photon detection correlations in this system, where the polarization state of one of the photons is deliberately modified. Inspired by Kwiat et al [7] and intrigued by what they found '*surprising*' when setting the phase angle of the wave function with the use of a quarter wave plate (QWP), we prepare a system of two photons in phase shifted entangled states by setting a QWP on one path to angles $\theta_{QWP}$ different than 0 or $\pi/2$, thereby phase-shifting part of the entangled photon states. To the best of our knowledge this is a novel approach when preparing phase shifted EPR-Bell states, as we have seen no reports in the literature of such experiments with various phase shifts, setting $\alpha$, as shown below in Eqn. (1), to any angle between 0 and $\pi$.

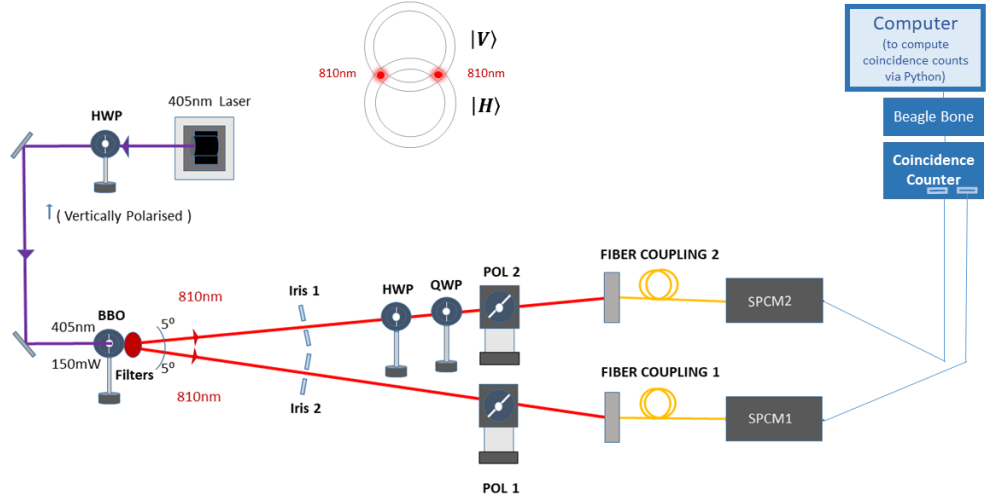

Fig. 1. Experimental design and setting to prepare EPR-Bell states using a Type II Beta Barium Borate down-conversion crystal (BBO) together with a 405 nm laser diode guided by mirrors towards the crystal. The half wave plate placed in one arm of the experiment together with a quarter wave plate are used to prepare the system in any EPR-Bell state, by choosing the correct angles. Polarizers 1 and 2 are used in conjunction with two Single Photon Counting Modules and a Coincidence Detector to compute the coincidence counts at different angles. The 810 nm pairs of photons are aligned with the aid of irises and an infrared camera to image the cones at the intersection. The upper cone is vertically polarized, while the lower is horizontally polarized.

We report the results of polarization correlation experiments with some intriguing questions about how to interpret them, with future possible application in quantum information, communication and computation. To start with we address the traditional EPR experiment, as shown in Fig. 1.

Denoting the polarization state of each photon $j$ with $j=1,2$, as $|H_j\rangle$ for horizontal and $|V_j\rangle$ for vertical, the general form of entangled states present at the intersection of the cones in Fig. 1 is described as follows:

$$|\psi\rangle = \frac{1}{\sqrt{2}}(|H_1, V_2\rangle + e^{i\alpha}|V_1, H_2\rangle),\qquad(1)$$

where $\alpha$ is determined by the birefringence of the crystal. By using additional birefringent elements, the four traditional EPR-Bell states can be created:

$$|\psi^\pm\rangle = \frac{1}{\sqrt{2}}(|H_1, V_2\rangle \pm |V_1, H_2\rangle)\qquad(2)$$

and

$$|\phi^\pm\rangle = \frac{1}{\sqrt{2}}(|H_1, H_2\rangle \pm |V_1, V_2\rangle).\qquad(3)$$

For instance, the sign in Eqn. (2) can be flipped by using a QWP angle $\theta_{QWP} = \pi/2$ in one of the paths. Similarly, one can go from $\psi$ to $\phi$ using a half-wave plate (HWP) under an angle of $\beta = \pi/4$ in one of the paths.

Beside the traditional EPR-Bell states we explore other states with different values for $\alpha$ that produce different phase shifts. In this paper, we study the effect of such phase shifts on the entangled states we call *Phase-Shifted EPR-Bell*.

In recent years some advances have been achieved in the study of quasi-Bell states [11] and semi-Bell states [12] and their applications to quantum information and quantum computation, like some cases of teleportation for example [13]. Both kinds of entangled states are different than the family of phase shifted states we investigate here.

The quasi-Bell states are defined as 'entangled qubit states based on a nonorthogonal computational basis' [14], while the semi-Bell states are defined as superpositions with unequal probability amplitudes of the component states in Eqns. (2) and (3) [12]. Our case study is different than both quasi- and semi-Bell states, since we keep our original EPR-Bell states derived from two orthogonal pure states $|H\rangle$ & $|V\rangle$, where we only modify the parameter $\alpha$ in Eqn. (1) to obtain such phase shifted entangled states.

In our measurements, we find phase shifted EPR-Bell states by setting the QWP to either $\theta_{QWP} = \pi/8$ or $\theta_{QWP} = \pi/4$, with the HWP plate to either $\beta = 0$ or $\beta = \pi/4$, and we adjust the angles in the computations of **S** when testing for violations of CHSH. We measure coincident photons detected in the two paths of the experiment.

It is important to note that a model has been introduced by Werbos and Fleury [15] hinting at the expected effects when setting a QWP at angles $\theta_{QWP}$ other than 0 or $\pi/2$ in one path in the Bell experiment. In their paper they explore models dealing with Time-Symmetric Quantum Measurement where the traditional "collapse of the wave function" is modelled as stochastic events that occur in optics, such as polarizers, representing different states into density matrices to detect a mix of possible outcomes. They also propose a Markov Quantum Electrodynamic model (MQED). They suggest that the MQED predictions can be calculated by a simple, local discrete Markov Random Field (MRF) formalism that can treat the effects of a QWP in one path of the Bell experiment and serve as a platform for the development of new technologies.

We chose the QM formalisms instead to compute our predictions and carry out our analysis, and we took into consideration the fact that they explored different angles for the QWP and the observed effects in their results. We explored a larger set of values for the parameter $\alpha$ and studied their implications further.

## 2. Method

### 2.1 Mathematical Background

We define $\theta_1$ and $\theta_2$ as the angles set for polarizers 1 and 2 respectively. In order to compute the theoretical coincidence counts function for each experimental setting, we use the quantum mechanics (QM) formalism [16] [17] from which we derive the landscape of theoretical values by changing the angles $\theta_1$ and $\theta_2$, according to the $\mathbf{C_T}(\theta_1, \theta_2)$ equations in Table 1. $\mathbf{C_T}(\theta_1, \theta_2)$ specifies the joint probability distribution of coincidence counts being detected at angles $\theta_1, \theta_2$.

To compare the theoretical results to the experimental ones, we also compute the normalized experimental coincidence counts $\mathbf{C_E}(\theta_1, \theta_2)$ obtained for the four EPR-Bell states, as shown in [7].

**Table 1**

| EPR-Bell states | $C_T(\theta_1, \theta_2)$ |
|---|---|
| $|\psi^+\rangle$ | $\sin^2(\theta_1 + \theta_2)$ |
| $|\psi^-\rangle$ | $\sin^2(\theta_1 - \theta_2)$ |
| $|\phi^+\rangle$ | $\cos^2(\theta_1 - \theta_2)$ |
| $|\phi^-\rangle$ | $\cos^2(\theta_1 + \theta_2)$ |

Shows EPR-Bell states with their corresponding equations for $C_T(\theta_1, \theta_2)$ theoretical.

Furthermore, we compute the **S** parameter to test for the Clauser, Horne, Shimony and Holt inequality (CHSH inequality) [18] by applying the following formulas, where we replace $C$ with $C_T$ or $C_E$ accordingly.

Defining: $E(\theta_1, \theta_2) = \frac{C(\theta_1,\theta_2)+C(\theta_1^\perp,\theta_2^\perp)-C(\theta_1,\theta_2^\perp)-C(\theta_1^\perp,\theta_2)}{C(\theta_1,\theta_2)+C(\theta_1^\perp,\theta_2^\perp)+C(\theta_1,\theta_2^\perp)+C(\theta_1^\perp,\theta_2)}$

(4)

where,

$$\mathbf{S} = E(\theta_1, \theta_2) - E(\theta_1', \theta_2) + E(\theta_1, \theta_2') + E(\theta_1', \theta_2')$$

(5)

Note that $\theta_1, \theta_1^\perp, \theta_1', \theta_1'^\perp, \theta_2, \theta_2^\perp, \theta_2', \theta_2'^\perp$ are angles chosen for maximum violations of the CHSH inequality, where $2 < |S| \leq 2\sqrt{2}$. In the CHSH inequality the number $2\sqrt{2}$ refers to Tsirelson's bound where strong violations approach $2\sqrt{2}$. Also note that the symbol $\perp$ means orthogonal, where for example, $\theta_1'$ is orthogonal to $\theta_1'^\perp$.

All the expectation value formulas $E(\theta_1, \theta_2), E(\theta_1', \theta_2), E(\theta_1, \theta_2'), E(\theta_1', \theta_2')$ to calculate S are provided in Appendix E.

If we define the operators $\widehat{P_{\theta_1}}$, $\widehat{P_{\theta_2}}$, $\widehat{Q_{\theta_{QWP}}}$, $\widehat{H_\beta}$, for polarizers 1 & 2, QWP and HWP respectively, then we can describe any phase shifted EPR-Bell State as:

$$\left|\psi_{\theta_{QWP},\beta}\right\rangle = \frac{1}{\sqrt{2}}(|H_1\rangle\widehat{Q_{\theta_{QWP}}}\widehat{H_\beta}|V_2\rangle + |V_1\rangle\widehat{Q_{\theta_{QWP}}}\widehat{H_\beta}|H_2\rangle)$$

(6)

In general, to obtain the prediction equations ($C_{TP}$) as will be shown in section 3.2, we find:

$$C_{TP} = \frac{1}{\sqrt{2}}(\widehat{P_{\theta_1}}|H_1\rangle\widehat{P_{\theta_2}}\widehat{Q_{\theta_{QWP}}}\widehat{H_\beta}|V_2\rangle + \widehat{P_{\theta_1}}|V_1\rangle\widehat{P_{\theta_2}}\widehat{Q_{\theta_{QWP}}}\widehat{H_\beta}|H_2\rangle)$$

(7)

where,

$$\widehat{P_\theta} = \begin{bmatrix} \cos^2\theta & \sin\theta\,\cos\theta \\ \sin\theta\,\cos\theta & \sin^2\theta \end{bmatrix}$$

(8)

and

$$\widehat{Q_{\theta_{QWP}}} = e^{\frac{-i\pi}{4}} \begin{bmatrix} \cos^2 \theta_{QWP} + i\sin^2 \theta_{QWP} & (1-i)\sin \theta_{QWP} \cos \theta_{QWP} \\ (1-i)\sin \theta_{QWP} \cos \theta_{QWP} & \sin^2 \theta_{QWP} + i\cos^2 \theta_{QWP} \end{bmatrix}$$
(9)

and

$$\widehat{H_{\beta}} = e^{\frac{-i\pi}{2}} \begin{bmatrix} \cos 2\beta & \sin 2\beta \\ \sin 2\beta & -\cos 2\beta \end{bmatrix}$$
(10)

Once the experimental design is set, we prepare the system in different phase shifted EPR-Bell states with the aid of a pair of wave plates (a quarter and half wave plate) to produce the four EPR-Bell states. We can achieve this by using a QWP in one of the two experimental paths of the down converted photons. We set $\theta_{QWP} = 0$ to produce a value of $\alpha = 0$, or $\theta_{QWP} = \pi/2$ to obtain $\alpha = \pi$, as in Eqns. (2) and (3), for example. Here we are interested in angles $0 < \theta_{QWP} < \pi/2$, which implies any arbitrary value of $0 < \alpha < \pi$. To obtain Eqn. (3), we use a HWP set to $\beta = \pi/4$ in the same path of and before the QWP.

## 2.2 Experimental Description

In this section we describe our general experimental setting, as shown previously in Fig. 1. We used a Beta Barium Borate Crystal (BBO) to produce pairs of down converted photons via Spontaneous Parametric Down Conversion (SPDC) Type II with the use of a Thorlabs 150 mW, CW Laser Diode, with centre wavelength $\lambda_0 = 405 \pm 5$ nm with results for the parameter S shown in Table 2. We also used a CNILaser 150 mW, CW Laser Diode, with centre wavelength $\lambda_0 = 405 \pm 5$ nm with results for parameter ŝ shown in Table 5. We align the equipment with the aid of an infrared camera that allows us to image the two light cones, one horizontally and the other vertically polarized, as depicted in Fig. A in Appendix A. We target the overlapping areas of the cones $|H\rangle$ and $|V\rangle$, as shown in Figs. 1 and B (a) in Appendix B, by placing two irises in each path, see Fig. B (e-i), in order to guaranty that we can couple the light in state $|\psi\rangle$ into the fibre coupling system attached to a pair of Single Photon Counting Modules (SPCM). We calibrate the different angles associated with the Beta Barium Borate crystal (BBO) that we use, until the optical phase matching condition is achieved [19] [20].

We use a SPCM-AQRH-14-FC from Excelitas Technologies with a wavelength range of 400 nm to 1060 nm, with a fibre coupled silicon avalanche photodiode. The peak photon detection efficiency is greater than 70% at 700 nm over a 180 μm diameter. When a photon is detected, a Transitor-Transitor Logic (TTL) level pulse is produced. The dead time is typically 20 ns, with a maximum of 40 ns. The dark counts are specified at a maximum of 100 c/s. Our detectors show a photon detection efficiency of about 47% at 810 nm, when considering the specifications provided by the manufacturers, were the typical values for 650 nm and 830 nm are 65% and 45% efficiency, respectively.

In the setup, alignment and calibration process we obtain a maximum on average of ~30000 c/s per channel with associated coincidence counts of ~700 c/s detected with an Ortek Coincidence Detector (model 414A) with less than 10 ns coincidence time window. The value of ~2.33% (700/30000) is in agreement with [21].

Once the setup is aligned and ready for experiments, we optimize for coincidence counts and run several experiments using two polarizers set at different angles on each path. This setting, aided by Python computer routines, allows for the computation of the **S** parameter to test for CHSH inequality violations.

## 3. Results

*3.1 Results for Experiment I*

By setting $\theta_1 = 0$ and varying $\theta_2$ in steps of $\pi/20$ we compute the number of coincidences for each pair of angles until we reach $\theta_2 = \pi$ and we normalize these results in order to compute the S parameter. Then we set $\theta_1 = \pi/20$ and we repeat the same procedure by varying $\theta_2$ until we reach $\pi$ again and normalize this new set of results. We continue until this procedure is repeated as many times as necessary for $\theta_1$ to reach $\pi$. Normalization of entangled coherent states with an arbitrary relative phase $\alpha$ has been treated by [22] for different types of coherent EPR-Bell states, like symmetric coherent, asymmetric coherent and Schrödinger's cat states.

When the procedure is complete, we end up with a matrix of coincidence counts as well as another matrix of normalized coincidence counts. The matrix of coincidence counts is used to plot its associated landscape of results with a surface plot, as shown in Fig. 2. Since our equipment is in units of degrees we display our figures accordingly for the sake of clarity and aesthetics.

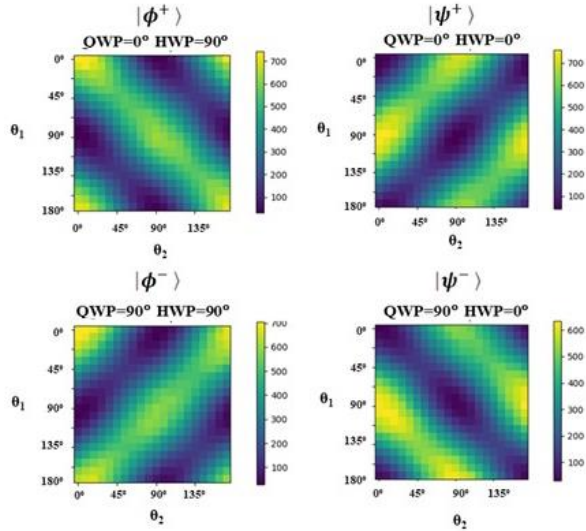

Fig. 2. Shows the landscapes for the experimental coincidence counts obtained for the EPR-Bell states. The y-axis gives the values for the different angles that $\boldsymbol{\theta_1}$ is set for polarizer 1, while the x-axis gives the values for the different angles that $\boldsymbol{\theta_2}$ is set for polarizer 2.

The values plotted in Fig. 2 are obtained by running the experiment for twenty runs for all pairs of angles $\theta_1$ and $\theta_2$. For each run, we measure the coincidence counts associated to each pair of angles, $C_\mathbf{E}(\theta_1, \theta_2)$, from which we compute the mean and standard deviations over all experimental runs. We use these mean coincidence counts to display experimental landscapes, as shown in Figs. 2 and 3. In Fig. 2 we can observe some attenuations in the diagonals of the landscapes.

These attenuations are caused due to the effects introduced by the QWP, when the polarizers are set at different angles, and can be removed via a simple normalization, as shown in Fig. 3, where we display theoretical $C_T(\theta_1, \theta_2)$ versus experimental coincidence counts $C_E(\theta_1, \theta_2)$, for the four EPR-Bell experiments. For clarity, in the results below the contrast is normalized to unity for each scan of angle $\theta_1$ at fixed $\theta_2$.

From this set of plots, we can clearly observe that the landscapes for $|\psi^+\rangle$ & $|\psi^-\rangle$ are isomorphic yet rotated by $\pi/2$, and similarly for the $|\phi^+\rangle$ & $|\phi^-\rangle$ EPR-Bell states.

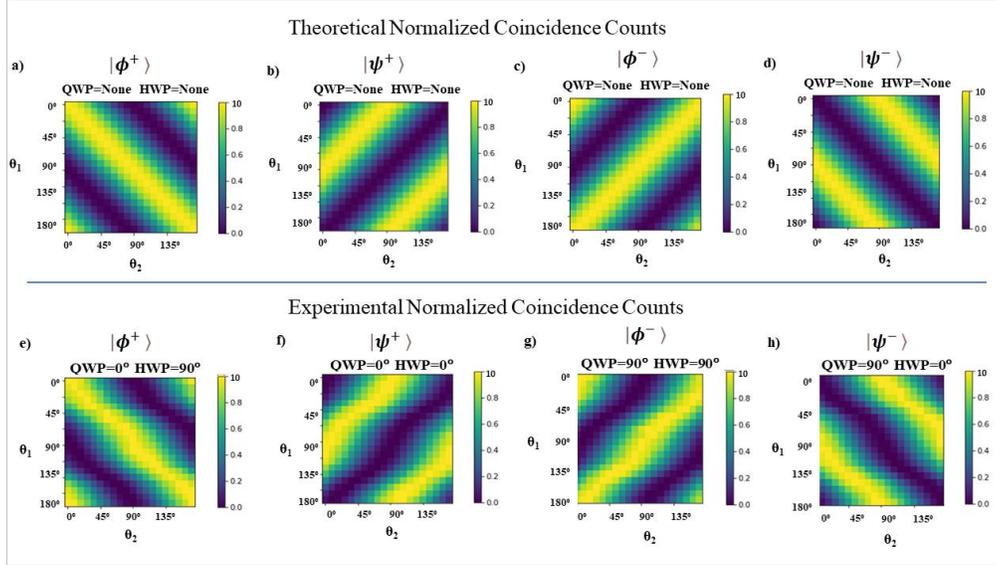

Fig. 3. Shows the landscapes for (a-d) theoretical coincidence counts versus (e-h) experimental normalized coincidence counts for the EPR-Bell states.

Finally, from Fig. 3 we can see that the normalized experimental results are qualitatively and quantitatively very similar to the theoretical results. All this qualitative analysis sets a preamble for exploring CHSH inequality violations by taking measurements on very specific angles and computing with them the **S** parameter, as we will show in the following section.

Choosing the appropriate values for $\theta_i$, $\theta_i^{(\perp)}$, $\theta_i'$, $\theta_1'^{\perp}$, our experimental results show strong violations of the CHSH inequality, where **S** closely approaches the Tsirelson's bound $S < 2\sqrt{2}$.

In Table 2 we summarize the results for each EPR-Bell state.

**Table 2**

| EPR-Bell states | S |
|---|---|
| $|\psi^+\rangle$ | -2.6840 ± 0.0064 |
| $|\psi^-\rangle$ | -2.7652 ± 0.0078 |
| $|\phi^+\rangle$ | 2.7812 ± 0.0070 |
| $|\phi^-\rangle$ | 2.6913 ± 0.0066 |

Shows the mean with confidence intervals for parameter S with very strong violations.

These results are in the same order of accuracy as those obtained by Kwiat et al [7], giving us an appropriate baseline to further test for *Phase-Shifted EPR-Bell states*.

## 3.2 Results for Experiment III

In this section we carry out both a qualitative and quantitative analysis, to understand the behaviour of phase shifted EPR-Bell states and their respective CHSH inequality violations.

In Fig. 4 we can clearly observe the effects of a shift in the original landscape of $|\psi^+\rangle$ by changing the angle $\theta_{QWP}$ of the QWP to $\frac{\pi}{8}$ and $\frac{\pi}{4}$, which prepares the system in what we call *Phase-Shifted EPR-Bell states* $|\psi^{+\frac{\pi}{8}}\rangle$ and $|\psi^{+\frac{\pi}{4}}\rangle$, respectively. The general form is $|\psi^{+\ \theta_{QWP}}\rangle$.

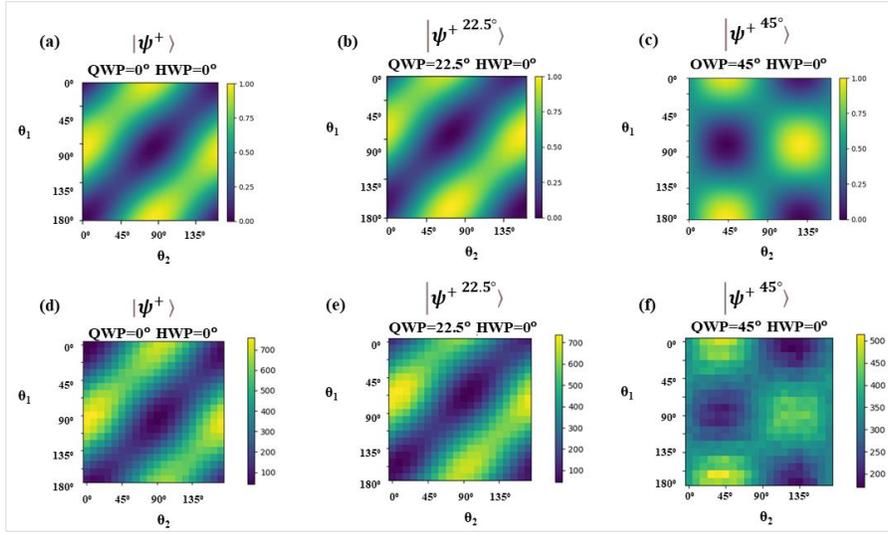

Fig. 4. Shows the phase shifted landscape for two different *Phase-Shifted EPR-Bell states*, with attenuation for both theoretical (top), for (a) $|\psi^+\rangle$, (b) $|\psi^{+\ 22.5°}\rangle$ and (c) $|\psi^{+\ 45°}\rangle$ and experimental results (bottom), where (d) is $|\psi^+\rangle$, (e) $|\psi^{+\ 22.5°}\rangle$ and (f) $|\psi^{+\ 45°}\rangle$, where the general form is $|\psi^{+\ \theta_{QWP}}\rangle$.

In order to properly model *Phase-Shifted EPR-Bell states* we computed a new set of equations based on the formalisms of quantum mechanics that model the effects of the QWP placed in one path of the experiment, as shown further below in Table 3. For our experiments we set the phase in the QWP and therefore $\theta_{QWP}$ to $\pi/8$ or $\pi/4$.

The equations for the *Phase-Shifted EPR-Bell states* studied are:

$$|\psi^+\rangle = \frac{1}{\sqrt{2}}(|H_1, V_2\rangle + e^{i0}|V_1, H_2\rangle)$$

(11)

$$|\psi^{+\frac{\pi}{8}}\rangle = \frac{1}{\sqrt{2}}(|H_1, V_2\rangle + e^{\frac{i\pi}{4}}|V_1, H_2\rangle)$$

(12)

$$|\psi^{+\frac{\pi}{4}}\rangle = \frac{1}{\sqrt{2}}(|H_1, V_2\rangle + e^{\frac{i\pi}{2}}|V_1, H_2\rangle)$$

(13)

This prepares the system in states, as shown in Eqns. (11), (12) and (13), for values of α equal to 0, $\pi/4$ and $\pi/2$ associated with angles $\theta_{QWP}$ of 0, $\pi/8$ and $\pi/4$ respectively, when setting the QWP. The theoretical equations that model the *Phase-Shifted EPR-Bell states* with attenuations, $\mathbf{C_{TP}(\theta_1, \theta_2, \gamma)}$, are shown in Table 3 for one of the four EPR-Bell states, both original and phase-shifted. The results for the rest of the equations associated with $|\psi^{-\theta_{QWP}}\rangle, |\phi^{+\theta_{QWP}}\rangle$ and $|\phi^{-\theta_{QWP}}\rangle$ are given in Table A in Appendix D.

**Table 3**

| EPR-Bell states | $\mathbf{C_{TP}(\theta_1, \theta_2, \gamma)}$ |
|---|---|
| $\|\psi^{+\theta_{QWP}}\rangle$ | $\left\| \begin{array}{c} (2i)\sin(2\gamma - \theta_1 + \theta_2) \\ +\sin(2\gamma + \theta_1 + \theta_2) \end{array} \right\|^2$ |

Shows the phase adjusted theoretical equations with attenuations, for the $|\psi^{+\theta_{QWP}}\rangle$ *Phase-Shifted EPR-Bell states* $C_{TP}(\theta_1, \theta_2, \gamma)$, where $\gamma$ is equal to $\theta_{QWP}/2$, and models the phase by which the EPR-Bell states are shifted via the QWP.

The equations obtained when the QWP was set at $\pi/4$, are shown in Table 4, and they produce a set of very unusual landscapes accompanied with no violations to the CHSH inequality with a predicted value for S = ±2. The landscapes of these EPR-Bell states resemble a classical behaviour, for both theoretical versus experimental landscapes, as seen in Fig. 4 (c and f) for $|\psi^{+\frac{\pi}{4}}\rangle$. We also note that a similar type of behaviour happens for angles $3\pi/4$, $5\pi/4$ and $7\pi/4$, $|\psi^{+\frac{3\pi}{4}}\rangle, |\psi^{+\frac{5\pi}{4}}\rangle$ and $|\psi^{+\frac{7\pi}{4}}\rangle$.

Following in Table 4, we present the equations that produce theoretical predictions and landscapes for $|\psi^{+\frac{\pi}{4}}\rangle$, when we set the QWP at $\pi/4$. The associated landscapes are shown in Appendix C in Fig. C, together with the rest of the equations for $|\psi^{+\frac{\pi}{4}}\rangle, |\psi^{-\frac{\pi}{4}}\rangle, |\psi^{+\frac{\pi}{4}}\rangle$ $|\phi^{+\frac{\pi}{4}}\rangle$ and $|\phi^{-\frac{\pi}{4}}\rangle$ in Table B in Appendix D.

**Table 4**

| EPR-Bell states | $\mathbf{C_{TP}(\theta_1, \theta_2, \gamma)}$ |
|---|---|
| $\|\psi^{+\frac{\pi}{4}}\rangle$ | $\left\| \begin{array}{c} \sin\left(2\frac{\pi}{8} + \theta_1 + \theta_2\right) \\ +(i)\sin(2\frac{\pi}{8} - \theta_1 + \theta_2) \end{array} \right\|^2$ |

Shows the phase adjusted theoretical equations with attenuations, for the *Phase-Shifted EPR-Bell state* $|\psi^{+\frac{\pi}{4}}\rangle$, $C_{TP}(\theta_1, \theta_2, \gamma = 2\frac{\pi}{8})$, where $\gamma$ is $\theta_{QWP}/2 = \frac{\pi}{8} \, rad$, and models the phase by which this EPR-Bell state is shifted.

To understand why Kwiat et al [7] found this surprising, we investigated the behaviour of the coincidence counts of the *Phase-Shifted EPR-Bell states*, and discovered that, at least for the angles studied, the QWP angle is half of $\gamma$, leading to $\alpha = 2\theta_{QWP} = 4\gamma$. This helps to establish the relationship between: (a) the physical angle $\theta_{QWP}$, set on the QWP, (b) the angle $\alpha$ in Eqn. (1), and (c) the angle $\gamma$, as was shown in the theoretical equations in Table 3. It took some further investigation to find the proper associations between the angles $\theta_{QWP}$ and $\gamma$ in the theoretical equations, since the angle $\alpha$, as expressed in [7], is "*surprisingly*" twice the angle set in the QWP.

In Fig. 5, we show theoretical values (top) and experimental results (bottom). Note that the theoretical values capture both the effect of the QWP and Polarizer, as observed in the phase and amplitude modulation. These are the kind of signals that produce the landscape in Fig. 4 (a) and (d), for $|\psi^+\rangle$.

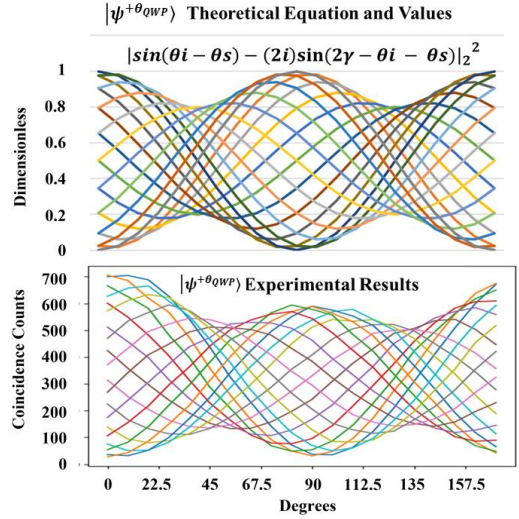

Fig. 5. Theoretical vs experimental results for $|\psi^{+\theta_{QWP}}\rangle$ based on formulas in Table 4, for $\gamma = 0$.

In Fig. 6 the reader is given a simplified version of Fig. 5 (top), showing around half of the curves in different colours, one curve per value of angle $\theta_1$ vs $\theta_2$ for values between 0 and $\pi$ in steps of $\pi/20$. This is intended to make it easier to understand the plots in Fig. 5.

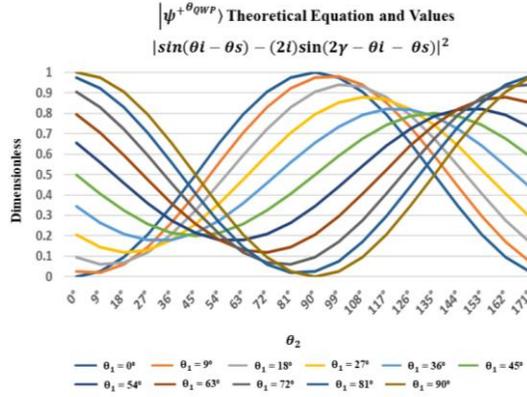

Fig. 6. Shows curves associated with different values of angle $\theta_1$ versus angle $\theta_2$ for values between 0º and 180º, in steps of 9º, for the theoretical values for $|\psi^{+\theta_{QWP}}\rangle$.

In Fig. 7, we display the theoretical values and experimental results for $|\psi^{+\frac{\pi}{8}}\rangle$, when the proper corrections for the $\pi/8$ phase shift are made. The corrected states $|\psi^{+\theta_{QWP}}\rangle^\Delta$ use the letter $\Delta$ as a superscript to refer to the Phase Adjusted Phase Shifted EPR-Bell states.

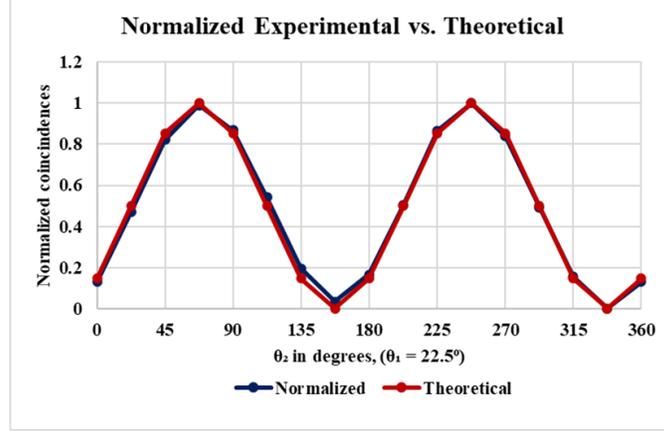

Fig. 7. Shows the adjusted values to compensate for the 22.5° shift when contrasting the normalized experimental results for $|\psi^{+\,22.5°}\rangle^\Delta$ with the theoretical values for $|\psi^+\rangle$.

The values for the violations computed as $\hat{S}$ are significant, as displayed in Table 5, apart from the values for $|\psi^{+\frac{\pi}{4}}\rangle^\Delta$.

The reader must note, however, that as $\theta_{QWP}$ shifts towards $\theta_{QWP} = \pi/4$, $\hat{S}$ tends to shift to values significantly below the Tsirelson's bound ($2\sqrt{2}, \sim 2.828$) and equal to 2 for $\theta_{QWP} = \pi/4$. This may need more theoretical investigation to verify experimental results.

Table 5

| Phase Adjusted - Phase Shifted EPR-Bell states | $\hat{S}$ |
|---|---|
| $|\psi^{+\,0}\rangle^\Delta$ | $-2.780 \pm 0.006$ |
| $|\psi^{+\,-\frac{\pi}{8}}\rangle^\Delta$ | $-2.795 \pm 0.009$ |
| $|\psi^{+\,\frac{\pi}{8}}\rangle^\Delta$ | $-2.778 \pm 0.008$ |
| $|\psi^{+\,\frac{\pi}{4}}\rangle^\Delta$ | $-2.025 \pm 0.011$ |

Shows the mean with confidence for parameter $\hat{S}$ for Phase Adjusted Results intervals, at 95% significance for a sample size of n = 500. The letter $\Delta$ as a superscript, refers to the Phase Adjusted Phase Shifted EPR-Bell states.

We also found values of S < 2 when we set $\theta_{QWP}$ to $\pi/4$, $3\pi/4$, $5\pi/4$ and $7\pi/4$.

We conduct similar tests and analysis for $|\psi^{-\theta_{QWP}}\rangle$, $|\phi^{+\theta_{QWP}}\rangle$, $|\phi^{-\theta_{QWP}}\rangle$ with different values for $\theta_{QWP}$ and obtain similar results. The landscapes can be found in Appendix B, in Fig. B.

## 4. Discussion, Conclusions and Future Perspectives

The experiments we conduct and the results we gather, match the results obtained by [7] for the four traditional EPR-Bell States $|\psi^\pm\rangle, |\phi^\pm\rangle$. However, the results that we obtain from the experiments on the *Phase-Shifted EPR-Bell states* $|\psi^{+\frac{\pi}{8}}\rangle$ and $|\psi^{+\frac{\pi}{4}}\rangle$ raise some questions about how to interpret these results.

Here we present some relevant observations:

(a) When $\theta_{QWP}$ is different than 0 and $\pi/2$ and $\alpha$, in Eqn. (1), is different than 0 or $\pi$, we observe a phase shift in the landscape produced by the correlated signals, as shown in Fig. 4.

(b) Using the **Phase Shifted EPR-Bell states**, as shown in Table 5, we show significant violations of the CHSH inequality, apart from $|\psi^{+\frac{\pi}{4}}\rangle^A$. In general, phase shifted states can still be considered EPR-Bell states. As for the state $|\psi^{+\frac{\pi}{4}}\rangle$, this needs further investigation since both predictions and results give a value of S < 2, something unusual and contrasting with the hallmark of what constitutes "entangled" states when verified by violations of the CHSH inequality, where $2 < |S| \leq 2\sqrt{2}$.

(c) In Fig. 8 we show the theoretical values for signals and landscapes associated with $|\psi^{+\frac{\pi}{4}}\rangle$ for example, where we observe that the signals are in phase and antiphase, producing the *"unusual"* values showing no violations of the CHSH inequality.

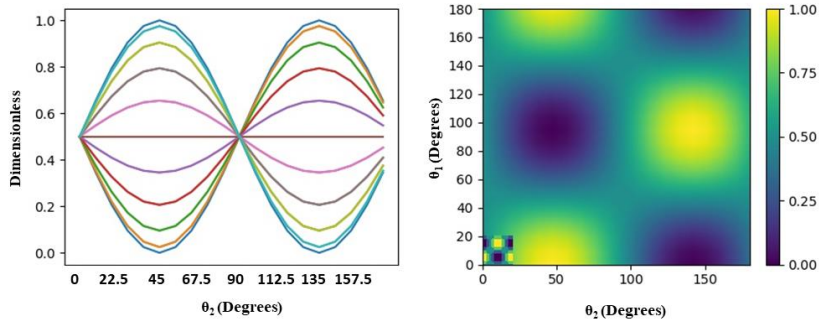

Fig. 8. Shows theoretical values for signals (left) and landscapes (right) associated with $|\psi^{+\ 45^0}\rangle$.

What is intriguing to us, is that the main determinant to obtain violations in our set up, is the angle at which we set the QWP. It is important to note that we replicated this experiment with two different brand-new **Zero-Order** QWPs to rule out the possibility of a QWP defect associated with the results obtained when setting the QWP at $\pi/4$, as previously shown in Fig. 4 (f).

(d) In general, we clearly see coincidence counts correlations as prescribed by quantum mechanics, yet the idea of entanglement described as "spooky action at a distance" is questionable given our results and the theoretical predictions for $\theta_{QWP}$ equal to $\pi/4$, $3\pi/4$, $5\pi/4$ and $7\pi/4$, since these are still correlated photons normally produced by the BBO.

In the future, we intend to carry out a more in-depth investigation and analysis, both theoretical and experimental, of *Phase-Shifted EPR-Bell states* at the boundaries of $\theta_{QWP}$ for angles $\pi/4$, $3\pi/4$, $5\pi/4$ and $7\pi/4$ both for $|\psi^{\pm \theta_{QWP}}\rangle$ and $|\phi^{\pm \theta_{QWP}}\rangle$. These *Phase-Shifted EPR-Bell states* may be useful for quantum information and quantum computation, and we conjecture that the parameter $\alpha$, when different than 0 or $\pi$, would produce a family of states characterized by such parameter and adding value to the field when addressing challenges like encryption and cyber security for example. This is clearly portrayed in [6] where phase shifts may happen due to weather conditions, which we have experimentally modelled, by dynamically changing $\theta_{QWP}(t)$, as displayed in Fig. 9.

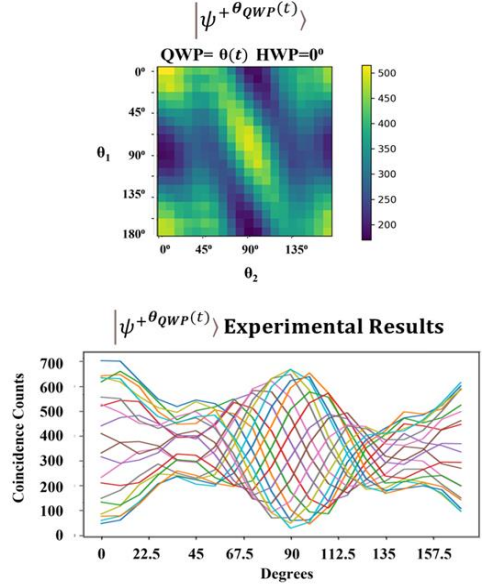

Fig. 9. Dynamical changes in phase $\boldsymbol{\theta_{QWP}(t)}$ for each change in polarization angle $\boldsymbol{\theta_1}$.

We look forward to explore other alternatives [23] [24] [25] [26] and to test the predictive power of these alternative models, while aiming to properly describe *Phase-Shifted EPR-Bell states*, particularly states like $|\psi^{+\frac{\pi}{4}}\rangle$, $|\psi^{+\frac{3\pi}{4}}\rangle$, $|\psi^{+\frac{5\pi}{4}}\rangle$ and $|\psi^{+\frac{7\pi}{4}}\rangle$.

In summary, we observe all the normal EPR-Bell states, but furthermore find phase-shifted versions of these states for various angles of the QWP (see Fig. 4) still showing significant violations of the CHSH inequalities. At an angle of $\pi/4$, however, our measurements show that the entanglement is lost, or hidden, and no violations are observed, something that resembles classical correlations. Therefore, we ask, why are these correlated photons showing no entanglement via the CHSH inequality with QWP at $\pi/4$? This remains a future challenge!

We leave the reader with a reflection about a quantum field theory (QFT) approach as described by Vitiello and Sabbadini [27], related to *"Phase-Mediated Entanglement in Quantum Field Theory"* where they express that *'phase-mediated long-range correlations over space-like distances find their origin in the coherent dynamical structure of the collective mode vacuum background. When entanglement is studied in the QFT formalism, in a natural way there are no "spooky forces at a distance" and violations of special relativity bound.'*

Based on our results and observations, careful attention and investigation of our unusual results is needed, given the historical and controversial nature, and still in the mind of many serious scholars, unfinished business of "entanglement" when understood as "spooky action at a distance". This suggests the need for a more moderate assertion based on experiments, instead of QM formalisms only, where we can assert that the photons are indeed correlated, yet "entanglement", understood as spooky action at a distance, remains inconclusive.

Concerning *Phase-Shifted EPR-Bell states* showing non-violations for the CHSH inequality, in the future, we intend to explore them with a robust Poincare-Block sphere approach, to derive a deeper understanding and potential explanation of this unusual phenomenon.

Finally, we can foresee that *Phase-Shifted EPR-Bell states* could play a role in quantum information, communication, and computation, by introducing a more diverse family of correlated states, as stated before.

**Acknowledgements:**

The authors consider Joshua Davis the main author of this paper and study, concerning experiments, analysis and the writing of the paper. Carey Jackman collaborated substantially with the experimental settings and the gathering and analysis of the results, with emphasis on the qualitative analysis. Rainer Leonhardt supported with experimental expertise and advice and in the reviewing and finalising of the paper. Paul Werbos, initially inspired a deeper investigation into the effects of a QWP setting on one arm of the entangled photon experiment and continuously gave further advice on how to address the effects of different settings of the QWP model, as described by Werbos and Fleury in [16]. Maarten Hoogerland, as the head of the lab, substantially oversaw the experimental procedure and the technical analysis of the results, actively participating in the refining of this document.

We would like to express our deep sense of gratitude to Kiko Galvez for always being ready to support with his experimental expertise, care and kindness. We are also grateful to Howard Carmichael and Giuseppe Vitiello for their heartfelt support and discussions about quantum mechanics, quantum field theory and entanglement in our meetings at different stages of the project.

**Disclosures:** The authors declare no conflicts of interest.

**Data availability:** Data underlying the results presented in this paper is not publicly available at this time but may be obtained from the authors upon reasonable request.

**Appendixes:**

**Appendix A - Vertical and Horizontal Cone Images:**

To align both paths with the 810 nm down converted photons, we used the infrared ProEM-HS Camera System from Princeton Instruments to image the cones produced by the Type II Beta Barium Borate (BBO) crystal, as shown in Fig. A:

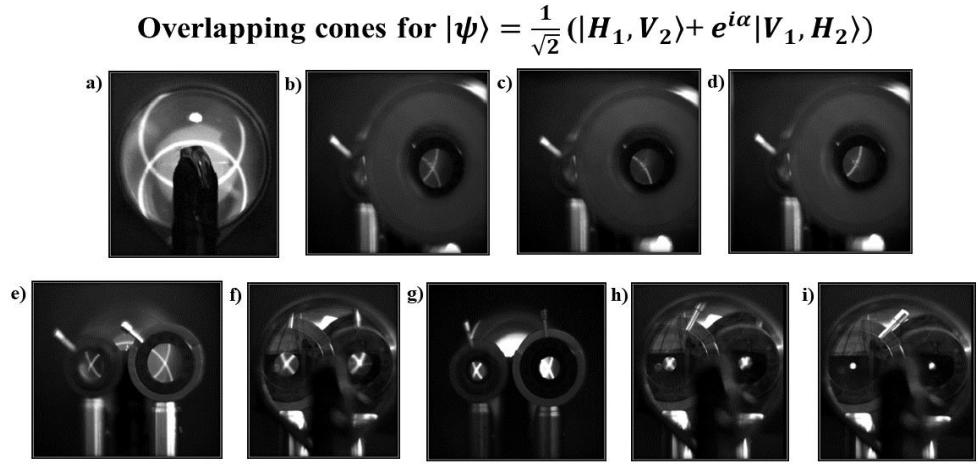

**Overlapping cones for** $|\psi\rangle = \frac{1}{\sqrt{2}}(|H_1, V_2\rangle + e^{i\alpha}|V_1, H_2\rangle)$

Fig. A. Shows in: (a) 810 nm light cones imaged by infrared camera, resulting from Spontaneous Parametric Down Conversion Type II, (b) sections of both vertical and horizontal cones through a polarizer, (c) Horizontal cone through the polarizer, (d) Vertical cone through the polarizer and (e-i) irises closed gradually to target the intersection of the two cones.

The bottom row shows a sequence where the irises are gradually closed until the overlapping areas of the cones $|H\rangle$ and $|V\rangle$ are targeted with precision. The top row shows the overlapping cones (top left), followed by a set of pictures clearly displaying each cone with horizontally and vertically polarized light, represented by $|H\rangle$ and $|V\rangle$ respectively. This is achieved by passing one of the intersections of both cones through a polarizer at $\pi/4$, $0$, and $\pi/2$, as shown in b), c) and d).

**Appendix B - Landscapes for the Phase-Shifted EPR-Bell States, $|\psi^{+\theta_{QWP}}\rangle$, $|\psi^{-\theta_{QWP}}\rangle, |\phi^{+\theta_{QWP}}\rangle, |\phi^{-\theta_{QWP}}\rangle$, for the QWP set at several angles:**

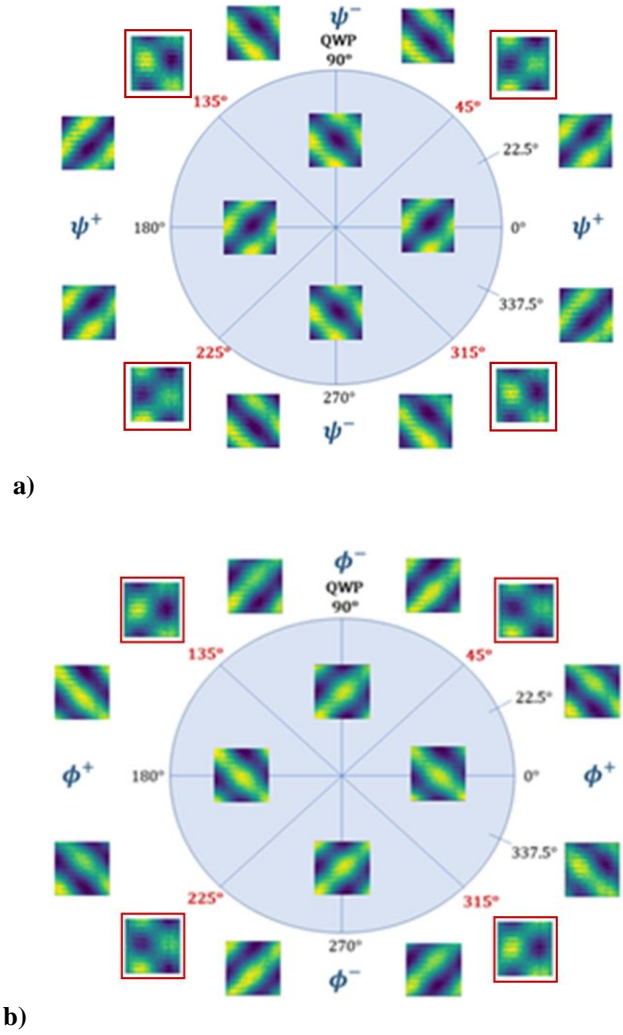

a)

b)

Fig. B a) $\boldsymbol{\psi}$ and b) $\boldsymbol{\phi}$. Shows the landscapes associated to angles set to produce violations of the CHSH inequality, such as: 0º, 22.5, 67.5º, 90º, 180º, 270º, 337.5º, as well as angles set at angles that produce no violations of the CHSH inequality, such as: 45º, 135º, 225º, 315º that are in red. Note that their features are different.

**Appendix C - Landscapes for the Phase-Shifted EPR-Bell States, $|\psi^{+45}\rangle, |\psi^{-45}\rangle, |\phi^{+45}\rangle, |\phi^{-45}\rangle$, for the QWP set at angles 45º producing very unusual and surprising classical landscapes where S < 2, showing no violations of the CHSH inequality:**

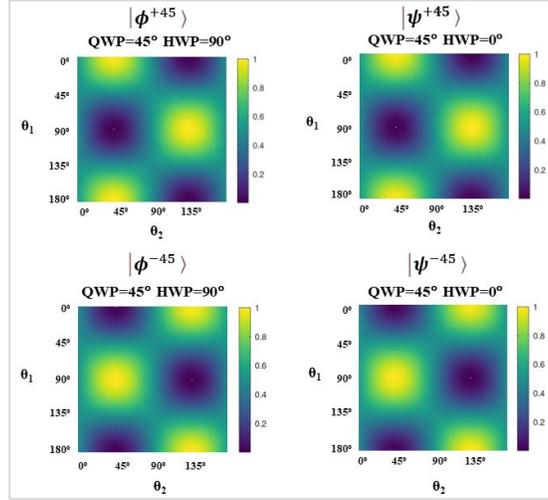

Fig. C. Shows the landscapes that produce no violations of the CHSH inequality when we set the QWP at 45º.

**Appendix D - Solution Equations for the QM formalisms:**

**Table A**

| EPR-Bell states | $C_{TP}(\theta_1, \theta_2, \gamma)$  (b) |
|---|---|
| $\|\psi^{+\theta_{QWP}}\rangle$ | $\left\| (2i)\sin(2\gamma - \theta_1 + \theta_2) + \sin(2\gamma + \theta_1 + \theta_2) \right\|^2$ |
| $\|\psi^{-\theta_{QWP}}\rangle$ | $\left\| (2i)\sin(-2\gamma + \theta_1 + \theta_2) - \sin(2\gamma + \theta_1 - \theta_2) \right\|^2$ |
| $\|\phi^{+\theta_{QWP}}\rangle$ | $\left\| (2i)\cos(-2\gamma + \theta_1 + \theta_2) + \cos(2\gamma + \theta_1 - \theta_2) \right\|^2$ |
| $\|\phi^{-\theta_{QWP}}\rangle$ | $\left\| (2i)\cos(2\gamma - \theta_1 + \theta_2) + \cos(2\gamma + \theta_1 + \theta_2) \right\|^2$ |

Shows the phase adjusted theoretical equations with attenuations, for the *Phase-Shifted EPR-Bell states* $C_{TP}(\theta_1, \theta_2, \gamma)$, where $\gamma$ is $\theta_{QWP}/2$, and models the phase by which the EPR-Bell states are shifted via the QWP.

**Table B**

| EPR-Bell states | $C_{TA}(\theta_1, \theta_2, \gamma)$   (a) |
|---|---|
| $\lvert\psi^{+\frac{\pi}{4}}\rangle$ | $\left\lvert \sin\left(2\frac{\pi}{8} + \theta_1 + \theta_2\right) + (i)\sin\left(2\frac{\pi}{8} - \theta_1 + \theta_2\right) \right\rvert^2$ |
| $\lvert\psi^{-\frac{\pi}{4}}\rangle$ | $\left\lvert \sin\left(2\frac{\pi}{8} + \theta_1 - \theta_2\right) + (i)\sin\left(-2\frac{\pi}{8} + \theta_1 + \theta_2\right) \right\rvert^2$ |
| $\lvert\phi^{+\frac{\pi}{4}}\rangle$ | $\left\lvert \cos\left(2\frac{\pi}{8} + \theta_1 - \theta_2\right) + (i)\cos\left(-2\frac{\pi}{8} + \theta_1 + \theta_2\right) \right\rvert^2$ |
| $\lvert\phi^{-\frac{\pi}{4}}\rangle$ | $\left\lvert \cos\left(2\frac{\pi}{8} + \theta_1 + \theta_2\right) + (i)\cos\left(2\frac{\pi}{8} - \theta_1 + \theta_2\right) \right\rvert^2$ |

Shows the phase adjusted theoretical equations with attenuations, for the *Phase-Shifted EPR-Bell state* $\lvert\psi^{+45°}\rangle$, $C_{TP}(\theta_1, \theta_2, \gamma = 2\frac{\pi}{8})$, where $\gamma$ is $\theta_{QWP}/2 = \frac{\pi}{8}\,rad$, and models the phase by which this EPR-Bell state is shifted.

## Appendix E - Expectation Value Formulas for the computation of S:

$$E(\theta_1, \theta_2) = \frac{C(\theta_1,\theta_2) + C(\theta_1^\perp,\theta_2^\perp) - C(\theta_1,\theta_2^\perp) - C(\theta_1^\perp,\theta_2)}{C(\theta_1,\theta_2) + C(\theta_1^\perp,\theta_2^\perp) + C(\theta_1,\theta_2^\perp) + C(\theta_1^\perp,\theta_2)} \tag{1}$$

$$E(\theta_1', \theta_2) = \frac{C(\theta_1',\theta_2) + C(\theta_1'^\perp,\theta_2^\perp) - C(\theta_1',\theta_2^\perp) - C(\theta_1'^\perp,\theta_2)}{C(\theta_1',\theta_2) + C(\theta_1'^\perp,\theta_2^\perp) + C(\theta_1',\theta_2^\perp) + C(\theta_1'^\perp,\theta_2)} \tag{2}$$

$$E(\theta_1, \theta_2') = \frac{C(\theta_1,\theta_2') + C(\theta_1^\perp,\theta_2'^\perp) - C(\theta_1,\theta_2'^\perp) - C(\theta_1^\perp,\theta_2')}{C(\theta_1,\theta_2') + C(\theta_1^\perp,\theta_2'^\perp) + C(\theta_1,\theta_2'^\perp) + C(\theta_1^\perp,\theta_2')} \tag{3}$$

$$E(\theta_1', \theta_2') = \frac{C(\theta_1',\theta_2') + C(\theta_1'^\perp,\theta_2'^\perp) - C(\theta_1',\theta_2'^\perp) - C(\theta_1'^\perp,\theta_2')}{C(\theta_1',\theta_2') + C(\theta_1'^\perp,\theta_2'^\perp) + C(\theta_1',\theta_2'^\perp) + C(\theta_1'^\perp,\theta_2')} \tag{4}$$